\documentclass{aa}
\usepackage{graphicx}
\usepackage{txfonts}
\usepackage[colorlinks=true,linkcolor=blue,urlcolor=blue, citecolor=blue]{hyperref}
\usepackage{enumitem}
\begin{document} 

   \title{Bridging the gap between Monte Carlo simulations and measurements of the LISA Pathfinder test-mass charging for LISA}

   

   \author{C. Grimani \inst{1,2}
          \and M. Villani \inst{1,2}
          \and M. Fabi\inst{1,2}
          \and A. Cesarini \inst{2}
          \and F. Sabbatini \inst{1,2}
          }

   \institute{DiSPeA, University of Urbino Carlo Bo, Urbino (PU), Italy\\
        \email{catia.grimani@uniurb.it}
    \and{INFN, Florence, Italy}
    }     
   \date{}

 
  \abstract
   {Cubic gold-platinum free-falling  test masses (TMs) constitute the mirrors of  future LISA and LISA-like interferometers for low-frequency gravitational wave detection in space. High-energy particles of Galactic and solar origin charge the TMs and thus induce spurious electrostatic and magnetic forces that limit the sensitivity of these interferometers.  Prelaunch Monte Carlo simulations of the TM charging were carried out for the LISA Pathfinder (LPF) mission, that was planned to test the LISA instrumentation. Measurements and simulations were compared during the mission operations. The measured net TM charging agreed with simulation estimates, while the charging noise was three to four times higher.}
   {We aim to bridge the gap between LPF TM charging noise simulations and observations.}
   {New Monte Carlo simulations of the  LPF TM charging due to both Galactic and solar particles were  carried out with the FLUKA/LEI toolkit. This allowed propagating low-energy electrons down to a few electronvolt.}
   {These improved FLUKA/LEI simulations agree with observations gathered  during the mission operations  within 
statistical and Monte Carlo errors. The charging noise induced by Galactic cosmic rays is about one thousand charges per second. This value increases 
to tens of thousands charges per second during solar energetic particle events. 
Similar results are expected for the LISA TM charging.}
   {}

\titlerunning{Lisa Pathfinder and LISA test-mass charging}
   \maketitle

\section{Introduction}\label{intro}
The LISA Pathfinder \cite[LPF;][]{antonucci2011, lisapf2, armano2016, armano18} was the European Space Agency demonstrator of the Laser Interferometer Space Antenna  (LISA), the first  interferometer  for gravitational wave detection 
in space in the frequency range  2$\times$10$^{-5}$-10$^{-1}$ Hz \citep{amaro2017}. The LPF was launched from the Kourou cosmodrome in French Guyana on December 4, 2015, and reached the Earth-Sun system Lagrange point L1 at the end of January 2016. The mission ended on July 18, 2017. The LPF spacecraft (S/C) consisted of one S/C carrying two cubic gold-platinum test masses (TMs) of approximately 2 kg mass, playing the role of mirrors of the interferometer. The aim of the LPF mission was to study the sources of  noise that are expected to limit the sensitivity of LISA. The most relevant sources of noise \citep{EleCast}
are: a) a frequency-independent Brownian noise due to the residual gas pressure in the region close to the TMs; b) an actuation noise, associated with the TMs that are kept at the center of an electrode housing by actuation electrodes. This noise is dominant below 1 mHz and is proportional to f$^{-1}$, where f represents the frequency here and below; c) an interferometer sensing noise relevant above 10 mHz and  proportional to f$^{2}$; d) a low-frequency fluctuation of the average stray electrostatic fields below 1 mHz, proportional to f$^{-1}$; e) laser radiation pressure that is mainly relevant below 1 mHz and f) Poissonian noise associated with the TM charging. This noise appears dominant below 1 mHz and is proportional to f$^{-1}$.
Finally, to these sources of noise, the noise due to the expected stochastic background of Galactic white dwarf binaries must be added. This is relevant around 1 mHz  \citep[confusion noise,][]{ruiter2010}.

 Spurious Coulomb and magnetic forces between the TMs and surrounding electrodes were predicted to originate from  the  charging  process of the TMs due to Galactic and solar particles with energies higher than 100 MeV/n. A detailed study of the noise associated with TM net and effective charging was carried out before the LPF launch for both LPF and LISA  \citep{shaul2005}. The measurement of the charging noise carried out with LPF appears in Figure 3.11 of \citet{EleCast}. Estimates and measurements appear to vary as a function of  frequency and range between 2$\times$10$^{-16}$ and 2$\times$10$^{-15}$ m s$^{-2}$Hz$^{-0.5}$ between 10$^{-5}$ Hz and 10$^{-4}$ Hz. 
In order to limit the intensity of the forces that increase with the charge deposited on the TMs, a periodic discharging with ultraviolet light beams illuminating the electrode housing was carried out on board the LPF S/C \citep{lampsUV}. An analogous discharging process will be considered for LISA \citep{inchau}.
                                                                                         
Prelaunch Monte Carlo simulations carried out to estimate the net and effective charging (charging noise) of the LPF TMs during the mission operations \citep{grim04,voc04, grim1, voc05,ara05,wass2005,grim15} were carried out with the FLUKA \citep{flukacern1,flukacern2} and 
Geant4 \citep{geant4_3,geant4_2,geant4_1} toolkits by considering the same satellite geometry. These two sets   of simulations agreed excellently when the same input particle fluxes were considered.
It is worthwhile to recall that the net charging is given by the algebraic sum of the charges deposited on the TMs, while positively and negatively charged particles contribute to the effective charging.
Measurements of the TM charging carried out with LPF in  April 2016 \citep{armano2017b} showed that the net charging agreed  with the simulations, while the effective charging appeared higher by a factor of 3 to 4. Several possibilities were explored to determine the origin of this mismatch. It was evaluated that any cause associated with the satellite geometry and incident particle fluxes would have generated a disagreement between observations and simulations common to the net and effective charging. Conversely, the experimental evidence suggested that  a large number of  particles  with the same charge were entering and escaping the TMs and thus  contributed to the charging noise
without significantly increasing the net charging. This scenario is consistent with a large number of  very low-energy electrons that are produced by primary and secondary particles escaping the surface of the TMs and surrounding electrodes. In the FLUKA and Geant4 versions available before the LPF mission launch, the propagation of electrons was limited to 1 keV and 250 eV, respectively. In the majority of applications, this low-energy limit in the electron propagation  does not affect the results of the simulations because the paths of these particles in dense material such as gold are very limited, they are on the order of microns at most. However, in the case of future space interferometers where the potential difference between electrodes and TMs is approximately one volt, the role of very low-energy electrons cannot be neglected. In this paper we report the results of recent improved simulations. A new Monte Carlo code was written in order to include the propagation of very low-energy electrons escaping the electrodes and the TMs. This dedicated Fortran 90 Monte Carlo tool called Low Energy Ionization \citep[LEI,][]{mattia,grimcqg21,teorico} was used in combination with FLUKA. Ionization energy losses, kinetic emission, and quantum backscattering are considered in LEI. The  net and effective charging for LPF in 2016 are re-estimated here. On the basis of lessons learned with the LPF,  reasonable expectations for the TM charging  induced by Galactic cosmic rays (GCRs) and solar particles for LISA TMs are also presented. In Sections \ref{sec2} and \ref{sec3}, the Galactic and solar particle environment for LPF and LISA are discussed. In Section \ref{sec4}, the net and effective TM charging simulations and measurements are presented. In Section \ref{sec5}, the FLUKA/LEI Monte Carlo tools are described. In Section \ref{sec6}, the Monte Carlo program uncertainties are discussed. Finally, in Section \ref{sec7}, the new simulation results are compared to LPF observations. 
\section{Galactic and solar particle fluxes}\label{sec2}
An average of 13.8 g cm$^{-2}$ of S/C and instrument material surrounded the LPF TMs. A similar amount of material is expected to be found on board the LISA TMs. This average material thickness sets the minimum energy of hadrons and electrons that contribute to the TM charging to 100 MeV/n and 20 MeV, respectively. The GCRs and solar energetic particles (SEPs) associated with gradual  events \citep{reames20arxiv} typically lie in this energy range. 
Both cosmic rays and SEPs consist approximately of 90\% protons, 8\% helium nuclei, 1\% heavy nuclei, and 1\% electrons, where the percentages are meant in particle numbers to the total number. 
The overall GCR flux was observed  to vary by
a factor of 4 in the inner heliosphere  during the last three solar cycles \citep{a&aub}. The cosmic-ray intensity presents  quasi 11-year and quasi 22-year periodicities that are associated with the solar activity and  the global solar magnetic field  (GSMF) polarity change.
The LPF satellite was sent into orbit during the declining phase of solar cycle  24, which was characterized by a positive polarity period of the GSMF. The same is expected for LISA, which is scheduled to be launched in 2035 near the maximum of solar cycle 26 \citep{singh19}.
 \cite{grim07} showed that during positive polarity periods, the energy spectra, $J(r,E,t)$, of cosmic
rays at a distance $r$ from the Sun  at a time $t$  are well represented by  the
 symmetric model in the force field approximation by \citet[G\&A;][]{glax68}. By assuming  time-independent interstellar cosmic-ray intensities $J(\infty,E+\Phi)$ and an energy loss parameter $\Phi,$ it is found that

\begin{equation}
\frac{J(r,E,t)}{E^2-E^2_0}=\frac{J(\infty,E+\Phi)}{(E+\Phi)^2-E^2_0},
\label{eq:1}
\end{equation}


\noindent where $E$ and $E_0$ represent  the particle total energy and rest mass, respectively.
For Z=1 particles with a rigidity (particle momentum  per unit charge) higher than  100 MV, the effect of the solar activity is completely defined by the
  solar modulation parameter $\phi$, which at these energies is equal to $\Phi$  \citep[][and references therein]{griele}. The proton and helium interstellar spectra adopted in this work are reported in \cite{burger2000} and \cite{Shikaze:2006je}, respectively.  The solar modulation parameter, estimated on the basis of the \cite{burger2000} proton interstellar spectrum, was reconstructed from ground-based cosmic-ray data.\footnote{\url{http://cosmicrays.oulu.fi/phi/Phi_mon.txt}} Unfortunately, no helium energy spectrum at the interstellar medium is reported in \cite{burger2000}.

\noindent The particle spectra obtained with the G\&A model were parameterized according to the following equation:
\begin{equation}\label{eq:flux}
F(E) = \frac{A}{(E+b)^\alpha} E^\beta \   {\rm Particles\ (m^2\ sr\ s\ GeV\ n^{-1})^{-1}},
\end{equation}

\noindent where the parameter $b$ allows us to modulate the spectra at low energies, and the parameters $\alpha$ and $\beta$ set the trend at high energies.
Finally, $A$ is the normalization constant.
The agreement of equation \ref{eq:flux} with the G\&A model was discussed in \citet{apj1} and references therein.

\begin{figure}
  \resizebox{\hsize}{!}{\includegraphics{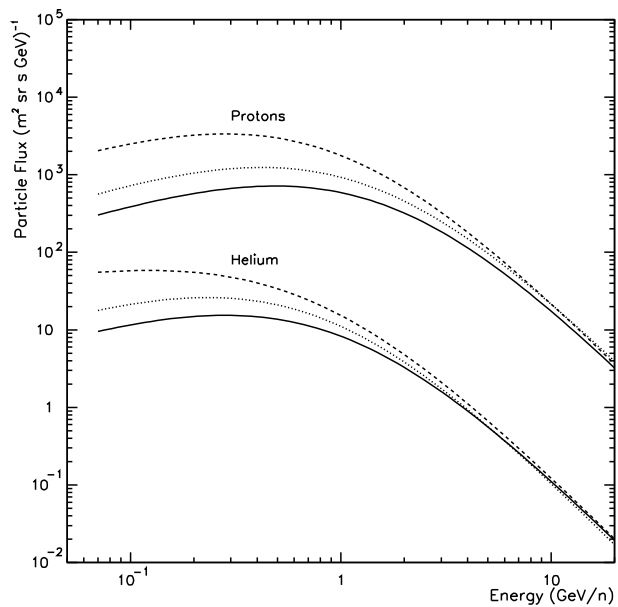}}
  \caption{Minimum (continuous lines) and maximum (dashed lines) proton  and helium  cosmic-ray energy spectra estimated for LPF 
before the mission launch (December 2015). The dotted lines indicate the cosmic-ray proton and helium energy spectra for BR 2492,
a few months after launch. 
The helium energy spectra have been scaled down by a factor of ten to avoid superposed lines.}
  \label{fig:p}
\end{figure}

Before the LPF launch, predictions of proton and helium
fluxes were carried out for the first part of the mission operations on the basis of the expected minimum and maximum solar activity for the period  corresponding to a solar 
modulation parameter ranging between 350 MV/c and 800 MV/c  \citep[details are reported in][]{grim15}. This estimate was carried out on the basis of  observations gathered in the past during
similar solar modulation conditions. In  Table \ref{tab:confronto} we report the  parameters $A$, $b$, $\alpha$ and $\beta$ 
that appear in equation \ref{eq:flux} for the LPF prelaunch predictions. The energy spectra are shown in Figure \ref{fig:p}. In the same table and figure, we also report the parameters and the particle 
spectra estimated for  Bartels rotation (BR) 2492 between March 31, 2016, and April 26, 2016, when the TM charging measurements were carried out in 
space. We recall that the BR number corresponds to the number of 27-day rotations of the Sun since
 February 8, 1832.

The  proton and helium fluxes estimated with the G\&A model  for BR 2492 when the solar modulation parameter $\phi$\footnote{\url{http://cosmicrays.oulu.fi/phi/Phi_mon.txt}} was 468 MV/c were compared  with  the AMS-02 experiment data gathered on the Space Station above 450 MeV/n and published in 2018 \citep{aguilar18} after the LPF mission 
was accomplished. The model-predicted proton flux appears to agree well with the data, while the helium flux is about 25\% higher. As a result, in Table \ref{tab:confronto}, Figure \ref{fig:p}, and in the  simulations, we adopted the helium flux normalized on the AMS-02 data. 
The different results of the model obtained for proton and helium energy spectra are ascribed to the helium flux at the interstellar medium reported in \citep{Shikaze:2006je}, which would require solar modulation parameters higher than those considered above. 

\begin{figure}
\centering
\resizebox{\hsize}{!}{\includegraphics{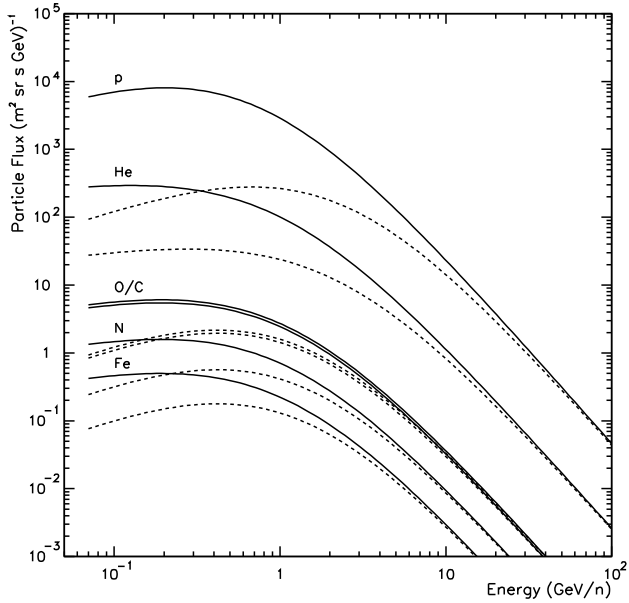}}
\caption{Cosmic-ray energy spectra at solar minimum ($\phi$= 200 MV/c;  continuous lines) 
and solar maximum ($\phi$= 1200 MV/c;  dashed lines). From top to bottom are reported the energy spectra of protons (p), helium (He), carbon (C), oxygen (O), nitrogen (N), and iron (Fe) nuclei.}\label{fig:ions}
\end{figure}

\begin{figure}
\centering
\resizebox{\hsize}{!}{\includegraphics{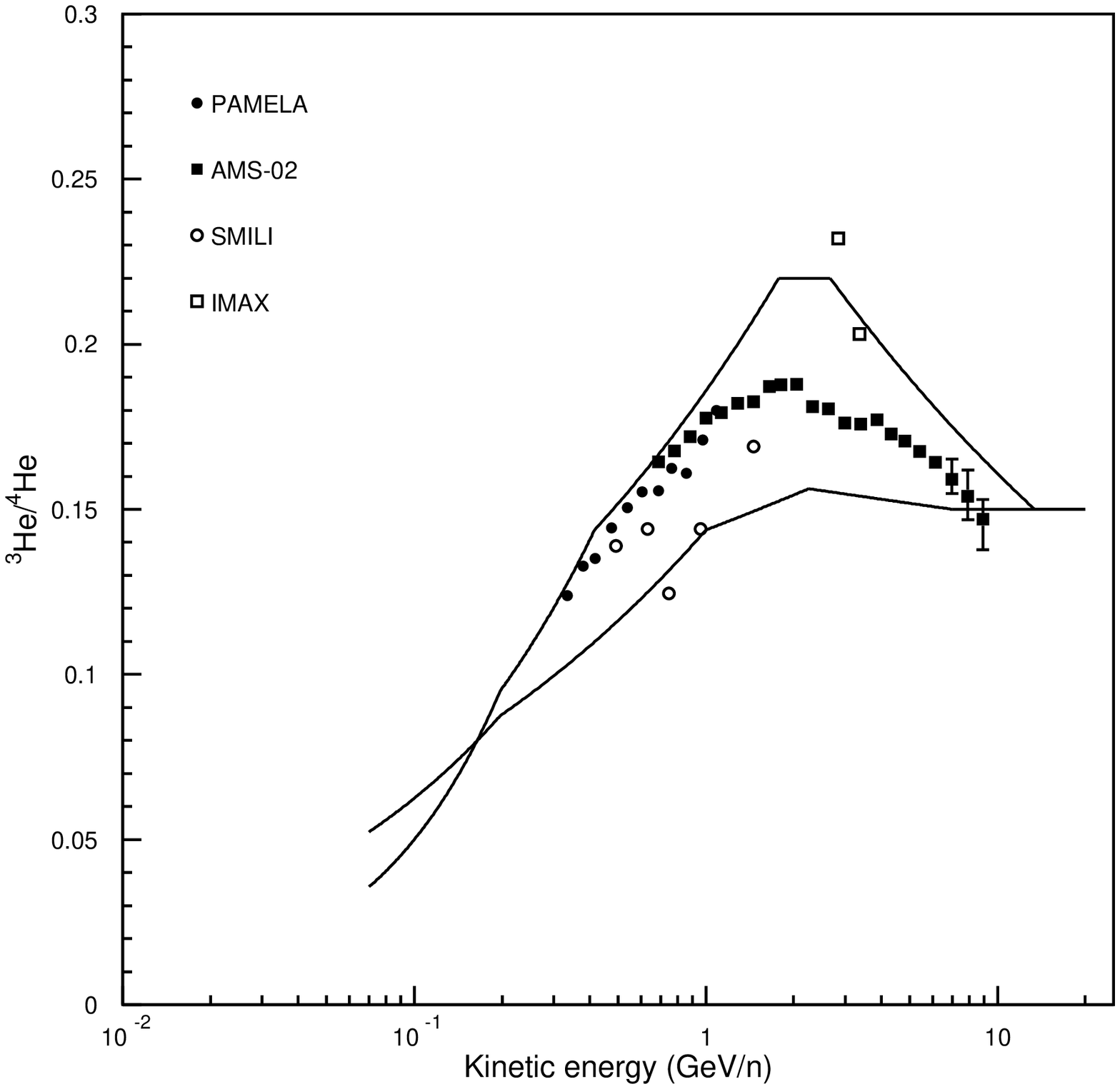}}
\caption{$^3He$/$^4He$ parameterization at solar minimum (top curve) and maximum (bottom curve). Data were gathered by the IMAX \citep{labrador03}, SMILI \citep{beatty93}, 
PAMELA \citep{adriani16}, and AMS-02 \citep{aguilar19} experiments.}\label{fig:he}
\end{figure}

\begin{figure}
\centering
\resizebox{\hsize}{!}{\includegraphics{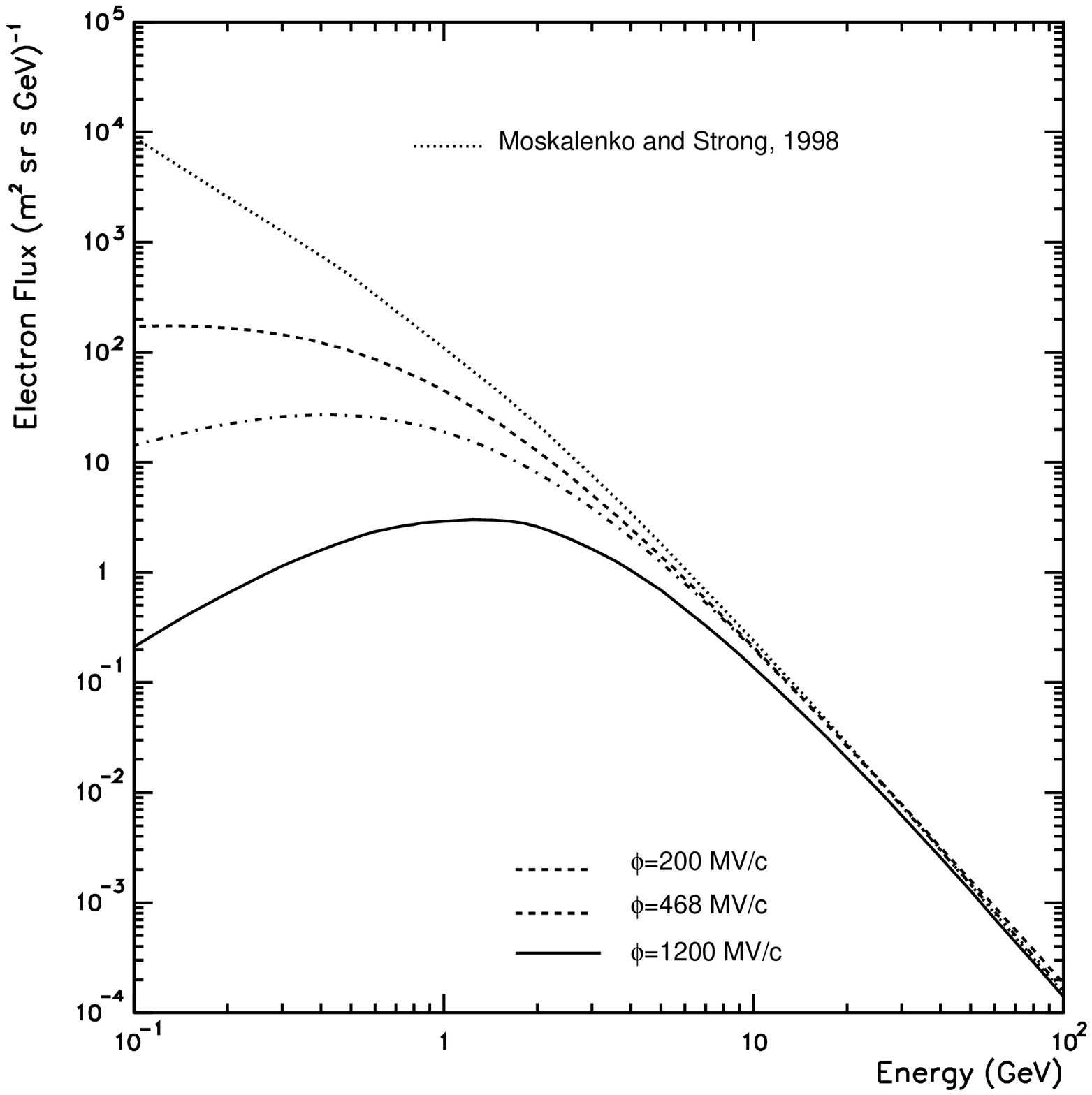}}
\caption{Galactic electron energy spectra at solar minimum (dashed line), solar maximum (continuous line) and during the BR 2492 (dot-dashed curve). The interstellar
spectrum is represented by the top dotted line \citep[][and references therein]{gri04, gri07}.}\label{fig:moska}
\end{figure}

In Tables \ref{tab:param} and \ref{tab:elepam} and Figures \ref{fig:ions}-\ref{fig:moska}, our predictions for proton, nucleus, and electron energy spectra for LISA are presented.
The LISA mission is supposed to be sent to space in 2035 at the maximum of solar cycle 26, for which the actual solar activity is unknown at present.
As a result, we considered a solar modulation parameter of $\phi$=200 MV/c at solar minimum and
$\phi$=1200 MV/c at solar maximum as extreme cases.

In order to disentangle the contribution of  $^3$He and  $^4$He isotopes from the overall He flux to the TM charging,
in Figure \ref{fig:he} we show the $^3$He/$^4$He ratio measured by the IMAX \citep{labrador03}, SMILI \citep{beatty93}, PAMELA \citep{adriani16}, and AMS-02 \citep{aguilar19} experiments. These observations were gathered at minimum and maximum solar modulation conditions and different GSMF polarity.
On the basis of the $^3$He/$^4$He data and models discussed in \cite{reimer98,ngobeni22}, we considered the parameterizations reported in Tables \ref{table:hemin} and \ref{table:hemax} and in Figure \ref{fig:he}.
Unfortunately, for the $^3$He/$^4$He ratio and the flux of nuclei with atomic number (Z) $>$2, no model can be considered reliable enough to estimate the particle energy spectra  during intermediate solar
modulation periods due to the limited amount of data gathered in space. 
In particular, model predictions introduce uncertainties larger than the contribution of these particles to the TM charging. As a result, for the simulations carried out for the BR 2492, we considered the overall helium flux as consisting of  $^4$He only, and the contribution of nuclei with Z$>$2 was estimated from solar minimum and maximum conditions \citep{papini96}.
These considerations do not apply to  electrons that have been measured by different experiments over several dozen years down to tens of MeV \citep[see for instance][]{gri04, gri07,pamele11,AMS02ele16}.

The Galactic and interplanetary electron contribution to the LPF TM charging was discussed in  detail in \cite{griele}.
The overall electron flux includes one Galactic component above approximately 100 MeV, one solar component below a few MeV associated with both impulsive and gradual
solar events, and one
Jovian component mainly below 20 MeV. Because particles below 20 MeV are not energetic enough to reach and charge the TMs, we considered only 
Galactic electrons. After proper modulation at 1 AU, the electron energy spectrum at the interstellar medium by \citet[][shown with the dotted line in Figure \ref{fig:moska},]{moska} was found to nicely represent observations gathered near Earth for different conditions of solar modulation and during different epochs of the GSMF \citep{gri04, gri07}. As a result, the electron spectra at 1 AU at solar minimum ($\phi$=200 MV/c; 
dashed line in Figure \ref{fig:moska}), at solar maximum ($\phi$=1200 MV/c; continuous line in Figure \ref{fig:moska}), and during BR 2492 ($\phi$=468 MV/c; dot-dashed line in Figure \ref{fig:moska}) were estimated for LISA and LPF with the G\&A model and the interstellar spectrum by \citet{moska}.
For the comparison of the simulated and measured LPF TM charging  during BR 2492, we have considered the proton, helium,  and electron energy spectra reported in Tables \ref{tab:confronto} and \ref{table:elepam2492}.
The preliminary  minimum and maximum estimates of the LISA TM net and effective charging were estimated by considering the LPF satellite geometry and solar minimum and maximum conditions for cosmic-ray protons, helium, carbon, nitrogen, oxygen, iron nuclei \citep{papini96}, and electrons \citep{griele}.


\begin{table*}
\renewcommand*{\arraystretch}{1.3}
\caption{LPF prelaunch minimum (top panel) and maximum (middle panel) cosmic-ray energy spectrum  predictions. The parameterizations were carried out according to the function F(E)=A(E+b)$^{-\alpha}$E$^\beta$ particles (m$^2$ sr s GeV n$^{-1}$)$^{-1}$. Estimates  of the cosmic-ray energy spectra in April 2016, when the TM charging measurements were carried out with LPF, are reported in the bottom panel.}              
\label{tab:confronto}      
\centering                                      
\begin{tabular}{c c c c c}         
\hline\hline                      
&\multicolumn{4}{c}{Minimum cosmic-ray flux  predictions for LPF ($\phi$=800 MV/c)}\\
\hline                                   
Particle & A & b &$\alpha$ & $\beta$\\
\hline
p  & 18000 & 1.54 & 3.67 & 0.88\\
He & 850 & 0.91 & 3.60 & 0.85 \\
\hline
&\multicolumn{4}{c}{Maximum cosmic-ray flux  predictions for LPF ($\phi$=350 MV/c)}\\
\hline
Particle & A & b &$\alpha$ & $\beta$\\
\hline
p  & 18000 & 0.88 & 3.68 & 0.89\\
He & 850 & 0.7 & 3.23 & 0.48\\
\hline
&\multicolumn{4}{c}{ Cosmic-ray flux during the LPF mission operations in April 2016 ($\phi$=468 MV/c)}\\
\hline
Particle & A & b &$\alpha$ & $\beta$\\
\hline
p  & 18000& 1.25 & 3.66 & 0.92\\
He &  850 & 0.74 & 3.68 & 0.85\\
\hline                                             
\end{tabular}
\end{table*}

\begin{table*}
\renewcommand*{\arraystretch}{1.3}
\caption{Parameterizations of proton and nucleus energy spectra at 1 AU at solar minimum ($\phi$=200 MV/c) and maximum ($\phi$=1200 MV/c).}              
\label{tab:param}      
\centering                                      
\begin{tabular}{ccccccccc}          
\hline\hline                        
& \multicolumn{4}{c}{Solar minimum} & \multicolumn{4}{c}{Solar maximum}\\
\hline
Particle & $A$ & $b$ & $\alpha$ & $\beta$& $A$ & $b$ & $\alpha$ & $\beta$\\
\hline
Protons  & 18000 & 0.65 & 3.66 & 0.87 & 18000 & 2.17 & 3.66 & 0.87\\
Helium   & 850   & 0.99 & 3.10 & 0.35 & 850   & 2.17 & 3.10 & 0.35\\
Carbon   & 28    & 1.05 & 3.25 & 0.50 & 28    & 1.15 & 3.75 & 1.00\\
Nitrogen & 7.3     & 1.05 & 3.25 & 0.50 & 7.3     & 1.15 & 3.75 & 1.00\\
Oxygen   & 25.2  & 1.05 & 3.25 & 0.50 & 25.2  & 1.15 & 3.75 & 1.00\\
Iron     & 2.3   & 1.05 & 3.25 & 0.50 & 2.3   & 1.15 & 3.75 & 1.00\\
\hline                                             
\end{tabular}
\end{table*}

\begin{table}
\renewcommand*{\arraystretch}{1.3}
\caption{Parameterization of electron energy spectra at solar minimum and solar maximum.}          
\label{tab:elepam}      
\centering                                      
\begin{tabular}{c c c c}         
\hline\hline                       
\multicolumn{2}{c}{Solar minimum}\\
Energy range & Parameterization \\
\hline
$>20$ MeV & $400 (E+0.82)^{-3.66} E^{0.5}$\\
\hline
\multicolumn{2}{c}{Solar maximum}\\
Energy range & Parameterization \\
\hline
50 MeV - 1 GeV & $4.5 (E -0.04)^{0.84}$\\
$>1$ GeV & $400 (E+2.5)^{-3.66} E^{0.5}$\\
\hline                                           
\end{tabular}
\end{table}

\begin{table}
\renewcommand*{\arraystretch}{1.3}
\caption{$^3$He/$^4$He ratio parameterization at solar minimum.}          
\label{table:hemin}      
\centering                                      
\begin{tabular}{c c}         
\hline\hline                       
Energy range (GeV/n) & $^3$He/$^4$He\\
\hline
0.07-0.197  & $0.44219$ $E^{0.94664}$\\
0.197-0.415 & $0.23439$ $E^{0.5559}$\\
0.415-1.778 & $0.1859$ $E^{0.29248}$\\
1.778-2.67  & $0.22$\\
2.67-7      & $0.277756$ $E^{-0.2374}$\\
7-13.40     & $0.277756$ $E^{-0.2374}$\\
$>13.40$    & $0.15$\\
\hline                                           
\end{tabular}
\end{table}

\begin{table}
\renewcommand*{\arraystretch}{1.3}
\caption{$^3$He/$^4$He ratio parameterization at solar maximum.}          
\label{table:hemax}      
\centering                                      
\begin{tabular}{c c}         
\hline\hline                       
Energy range (GeV/n) & $^3$He/$^4$He\\
\hline
0.07-0.197  & $0.195943\, E^{0.49625}$\\
0.197-1.    & $0.14375\, E^{0.3056}$\\
1.-2.254    & $0.14375\, E^{0.102597}$\\
2.254-7.    & $0.16089\, E^{-0.036}$\\
$>7$        & $0.15$\\
\hline                                           
\end{tabular}
\end{table}

\begin{table}
\renewcommand*{\arraystretch}{1.3}
\caption{Same as Table \ref{tab:elepam} for Galactic electrons during BR 2492.}          
\label{table:elepam2492}      
\centering                                      
\begin{tabular}{c c c c}         
\hline\hline                       
$A$ & $b$ & $\alpha $ & $\beta$\\
\hline
670 & 1.25 & 4.4 & 1.1\\
\hline                                           
\end{tabular}
\end{table}

\section{Solar energetic particle events during LISA}\label{sec3}

\begin{figure}
\centering
\resizebox{\hsize}{!}{\includegraphics{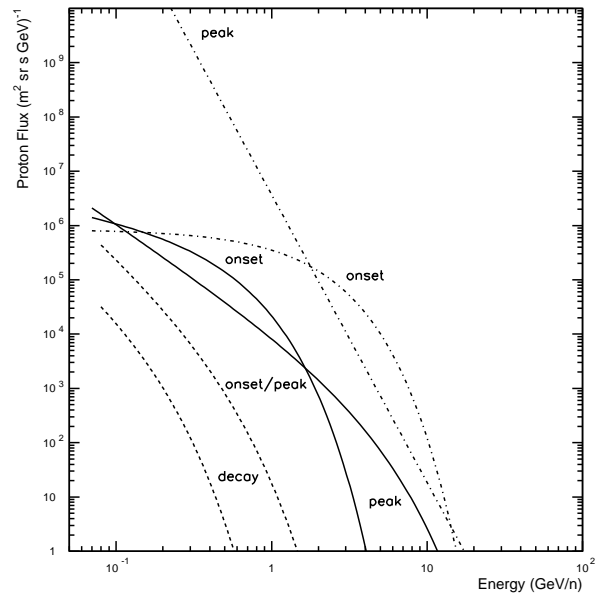}}
\caption{Solar energetic proton fluxes observed during the evolution of the gradual events dated February 23, 1956 (dot-dashed line), December 13, 2006  (continuous line), and December 14, 2006 (dashed line). The onset, peak, and decay phases of each event are indicated.}\label{fig:sep}
\end{figure}

Solar particle fluxes evolve in space, energy, and time during the events.
The parameterization of the solar particle energy spectra during different events  was discussed, for instance, in \cite{grimani2013}. 
In the majority of cases, the SEP energy spectra, at the onset of the events, show a power-law trend with an exponential cutoff, while at the peak, they
 appear compatible with a power-law function. Moreover, at the onset of SEP events, electrons are observed first due to particle velocity dispersion,
 and the particle spatial distribution  is characterized by small pitch angles with respect to the interplanetary magnetic field lines for magnetically well-connected events. Conversely, during the declining phase of the events, the spatial particle distribution becomes isotropic.
Impulsive SEP events are associated with proton acceleration below 50 MeV \citep{reameslibro} and play no role in the TM charging.
Unfortunately, during the LPF operations, no gradual SEP events were observed above the GCR background. We have considered here the LPF TM charging at the onset and at the peak of gradual SEP events of different intensities observed on February 23, 1956 \citep{vashe}, December 13, 2006,  and December 14, 2006 \citep{pamFD}, in order to obtain reasonable TM charging estimates for LISA.
The  proton fluxes observed during the evolution of these events are shown in Figure \ref{fig:sep}. The 2006 events were observed in space by the PAMELA
experiment, and the proton differential fluxes of the February 23, 1956, event were inferred from neutron
monitor data. The probability of  intense events such as the one on February 23, 1956, characterized by a proton fluence of 10$^9$ protons cm$^{-2}$ above 30 MeV, is one event 
every 60 years \citep{miro14}. This event  is considered here as a worst case for LISA. The fluence of most of the gradual SEP events ranges 
between 10$^6$ and 10$^7$ protons cm$^{-2}$ above 30 MeV. The SEP events with fluences of 10$^5$-10$^6$ protons cm$^{-2}$ are not observed at solar minimum
above the background of GCRs above 70 MeV. On the basis of observations gathered during previous solar cycles\footnote{\url{https://wwwbis.sidc.be/silso/datafiles}} and predictions for the next two cycles \citep{singh19}, the expected number of SEP events  during the LISA operations will range between 10 and 20 per year at most during the first part of the mission because the LISA launch is scheduled  at the maximum  of solar cycle 26.     


\section{LISA Pathfinder net and effective TM charging}\label{sec4}

The net ($\lambda_{net}$) and effective ($\lambda_{eff}$) charging rates of the LPF TMs are defined below. They are

\begin{equation}
\lambda_{net}=\sum^{+\infty}_{j=-\infty}\ j \lambda_j \ \ \ {\rm s^{-1}}
\label{eq:3}
\end{equation}

\begin{equation}
\lambda_{eff}=\sum^{+\infty}_{j=-\infty}\ j^2 \lambda_j \ \ \ {\rm s^{-1}},
\label{eq:4}
\end{equation}

\noindent where $j$ represents the net number of positive and negative charges deposited by single events, and $\lambda_{j}$ is the rate of occurrence of these events.
As we pointed out above, positive and negative charges cancel out in the net charging computation, while both positive and negative net deposited charges contribute to the effective charging. The spectral density of the charging shot noise (S) is expressed in terms of effective charging rate,
\begin{equation}
    S = \sqrt{2e^{2}\lambda_{eff}} \quad {\rm e\ s^{-1}\ Hz^{-0.5}},
\end{equation}

\noindent where e is the elementary charge.

\subsection{Lisa Pathfinder  prelaunch  TM charging  Monte Carlo simulations}

Monte Carlo simulations of the LPF TM charging were formerly carried out  with Geant4 and FLUKA for different conditions of solar modulation 
several years before the mission launch \citep{grim04,voc04, grim1, voc05,ara05,wass2005}. New Monte Carlo simulations based on FLUKA were performed in 2015 just before mission launch by considering as input cosmic-ray fluxes the fluxes that were estimated on the basis of the expected minimum and maximum solar activity at  the end of 2015 and beginning of 2016 presented in Section \ref{sec2} \citep{grim15}.
The TM charging obtained with these input fluxes  appears in Table \ref{tab:2015}. 
FLUKA and Geant4 simulations returned very similar results at solar minimum, as discussed in detail in \cite{grim15}, even though the energy limit for particle propagation in the two simulation codes was different and limited  for electrons, positrons, and photons to 1 keV in FLUKA and  to 250 eV in Geant4. 
This evidence was ascribed to the average ionization potential in gold of 790 eV \citep{mattia}, to which the secondary electron production and propagation 
was limited \textit{\textup{de facto}} in Geant4.  

\begin{table*}
\renewcommand*{\arraystretch}{1.3}
\caption{LPF prelaunch minimum (left) and maximum (right) TM charging predictions for the first part of the mission \citep{grim15}.}
\label{tab:2015}          
\centering                                      
\begin{tabular}{ccc|cc}         
\hline\hline                       
Particle & Net charging & Effective charging rate & Net charging & Effective charging rate\\
& (e$^{+}$ s$^{-1}$)  & (s$^{-1}$) & (e$^{+}$ s$^{-1}$) & (s$^{-1}$)\\
\hline
Protons & 14.1 & 168.9 & 32.5 & 295.5 \\
$^3$He & 0.22 & 0.92 & 1.9 & 5.6\\
$^4$He & 0.81 & 1.9 & 3.8 & 10.7\\
\hline                                           
\end{tabular}
\end{table*}

\subsection{LISA Pathfinder TM charging during mission operations}

The TM net and effective charging were measured on board the LPF satellite on April 20-23, 2016. 
The results are reported in Table \ref{tab:meas} \citep{armano2017b}.

\begin{table}
\renewcommand*{\arraystretch}{1.3}
\caption{Net and effective TM charging measured with LPF on April 20-23, 2016 \citep{armano2017b}.}\label{tab:meas}
\centering                                      
\begin{tabular}{c|cc}         
\hline\hline                       
& Test mass 1 & Test mass 2\\ 
\hline  
Net charging (e$^{+}$ s$^{-1}$) & $+22.9\pm 1.7$ & $+24.5\pm 2.1$\\
Effective charging rate (s$^{-1}$) & $1060\pm 90$ & $1360\pm 130$\\
\hline
\end{tabular}
\end{table}

Comparison of Tables \ref{tab:2015} and \ref{tab:meas} show that the measured net charging is in the middle of the prediction range, while the effective charging appears three to four times higher than expected. Any possible cause for the mismatch had to be plausibly associated with particles with the same charge sign entering and escaping the TMs, thus contributing to the noise without contributing to the net charging. 
Electrons and positrons, as the lowest-mass charged particles, propagate at keV energies over typical pathlengths of tens of microns in  solid materials.
When the propagation of these particles is considered below 1 keV,  ionization energy losses, multiple scattering, and quantum backscattering must be considered properly. As a result,  the computation  time  strongly increases, and consequently, it is in general neglected. In the majority of applications, the limited propagation lengths of these particles at low energies do not impact the results. Unfortunately, this is not the case of the LPF and LISA missions because the potential difference between the TMs and surrounding electrodes is about one volt and low-energy electrons strongly affect the TM charging process. The low-energy electron propagation  in the Geant4 toolkit currently depends on the libraries  adopted in the simulations.  In particular,  the electron production and propagation is limited to 100 eV \citep{ivan17} in Geant4/opt 4. 

\section{FLUKA/LEI tool}\label{sec5}
\begin{figure}
\centering

\resizebox{0.7\hsize}{!}{\includegraphics{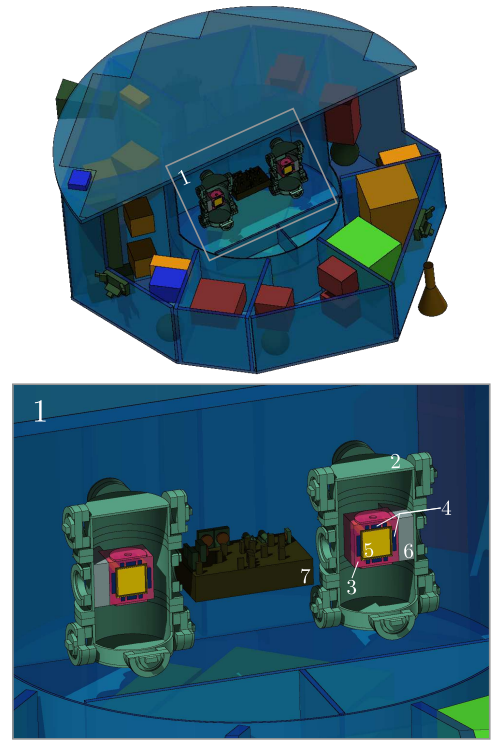}}
\caption{LPF combinatorial geometry model built with Flair for FLUKA: 1 LISA technology package, 2 vacuum enclosure, 3 electrode housing, 4 electrodes, 5 test mass, 6 gravitational compensation mass, and 7 optical bench. Items 2 to 7 are found around both TMs.}\label{fig:lpf_model}
\end{figure}

The LPF geometry has been built in FLUKA with the Flair interface \citep[][see Fig. \ref{fig:lpf_model}]{flair}.
The electron and positron production and propagation from keV energies down to the limit of their quantum  wave-like behavior has been taken into account in a new  Monte Carlo program written in Fortran 90,
LEI,
in order to include  the effects of very low-energy electromagnetic processes in the LPF TM charging simulations.  
The LEI Monte Carlo has been activated in the outer 150 nm of the gold-plated layers of the TMs and electrodes.
The thickness of these layers was chosen as a compromise between increasing computing time and the aim to include  all the electrons and positrons  contributing  to the TM net and effective charging in the simulations. In our simulation architecture,  incoming primary and secondary particles are propagated with FLUKA down to the 150 nm gold layers. All charged particles incident on the gold layers  constitute the input data for the Monte Carlo LEI, which allows us to simulate electron and positron low-energy electromagnetic processes  down to 12 eV when  the  diffraction is also activated \citep[see][for details]{mattia,grimcqg21,teorico}. The low-energy electromagnetic processes included in the LEI Monte Carlo are discussed below. 

Particles incident on the electrode and TM gold-plated layers are propagated by considering one  nanometer step, while the  electrons produced within 
the slab are propagated  
at one angstrom step. The program code is available upon request. 
When the FLUKA/LEI simulations are complete, all charged particles deposited into the TMs by each incident particle are counted and the net and effective charging are estimated.

\subsection{Low-energy particle ionization in LEI}
When charged particles propagate through matter, the target  material atoms are ionized. 
Low-energy ionization in the range 12-1000 eV was implemented in LEI. For all particles different from electrons (ions, pions, muons, etc.), we 
adopted the formula by \cite{cuci} to calculate the  number of electrons produced per kinetic energy interval, 
\begin{equation}
\frac{dn}{dK} = \frac{2\pi N e^4}{m c^2\beta^2} \frac{Z_s^2}{K^2} \left[ 1- \frac{\beta^2K}{E_m} + \frac{\pi\beta Z_s^2}{137} \sqrt{\frac{K}{E_m}} \left( 1- \frac{K}{E_m} \right) \right] dx,
\label{eq:5}
\end{equation}
where $N$ is the material electron density, $\beta$ is the incoming particle velocity, $e$, $m$, and $K$ are the emitted electron charge, mass, and kinetic energy, and $dx$ is the material thickness traversed by the incident particle. The maximum energy that can be transferred to an emitted electron is
\begin{equation}
E_m= \frac{2mc^2\beta^2}{1-\beta^2},
\label{eq:6}
\end{equation}
while the effective charge of the particle inside the material is given by
\begin{equation}
Z_s=Z\left[ 1- \exp\left( - \frac{125\beta}{Z^{2/3}} \right) \right], 
\label{eq:7}
\end{equation}
\noindent where $Z$ is the charge of the incident particle.
The ionized electron direction emission is finally estimated as follows:
\begin{equation}
\cos\theta= \sqrt{\frac{K}{E_m}},
\label{eq:8}
\end{equation}

 \noindent where $\theta$ is the angle between the direction of the ionizing particle and that of the emitted electron.

For electron-induced ionization, we adopted the  cross section reported below \citep{sakata},
\begin{eqnarray} \nonumber
\sigma&= \frac{4\pi a_0^2\alpha^4N_s}{\left( \beta_t^2+ \left( \beta_u^2+\beta_b^2 \right)/v \right)2b^\prime} \Bigg[ \frac{1}{2} \left(\ln\left( \frac{\beta_t^2}{1-\beta_t^2} \right)  -\beta_t^2 - \ln(2b^\prime)\right) \left( 1- \frac{1}{t^2} \right) +\\
&+ 1 - \frac{1}{t} - \frac{\ln(t)}{t+1} \frac{1+2t^\prime}{(1+t^\prime/2)^2} + \frac{b^{\prime 2}}{(1+t^\prime/2)^2} \frac{t-1}{2} \Bigg].
\end{eqnarray}
In the above formula, 
\begin{equation}
\beta_t^2=1-\frac{1}{(1+t^\prime)^2}  \qquad t^\prime = \frac{K}{mc^2},
\label{eq:10}
\end{equation}
\begin{equation}
\beta_u^2=1-\frac{1}{(1+u^\prime)^2}  \qquad u^\prime = \frac{U}{mc^2},
\label{eq:11}
\end{equation}
\begin{equation}
\beta_b^2=1-\frac{1}{(1+b^\prime)^2}  \qquad b^\prime = \frac{B}{mc^2},
\label{eq:12}
\end{equation}
\begin{equation}
t=\frac{K}{B},
\label{eq:13}
\end{equation}
where $K$ is the incident electron kinetic energy, $U$ is the bound kinetic energy of the electron inside the atom, $B$ is the bound electron binding energy, $N_s$ is the occupation number of the shell to be ionized, and $\alpha$ is the fine structure constant. Finally, $v$ is an empirical parameter  set equal to the  principal quantum number of the atomic shell to be ionized.

Ionization contributes most to the low-energy electron production  \citep{mattia}. 
Kinetic emission and quantum backscattering balance out the effect of ionization, leading to a positive net charge of the TMs while increasing the effective charging.

\subsection{Kinetic emission}
Electron emission  may follow when a particle below keV energies crosses a material slab.  
If the incoming particle is an ion, this process is called ion-induced electron emission (IIEE), while if it is an electron it is called electron-induced electron emission (EIEE): we refer to both  processes as kinetic emission. 

Kinetic emission is the macroscopic effect of many microscopic processes such as plasmon production and decay and both elastic and inelastic scattering \citep{grimcqg21}. The overall electron production is represented by the yield, that is, by the number of electrons emitted per incoming particle. In \cite{grimcqg21}, we calculated the expected electron yield for IIEE and EIEE from gold by using the Schou approach \citep{schou80}. We found  that for EIEE, the peak of electron emission is at about  100 eV, while for IIEE,  the maximum emission occurs at higher energies for an increasing  atomic mass number $A$ of the incident ion, and also that this maximum increases with the ion charge $Z$.

The kinetic emission process has been implemented in LEI using the yield: Whenever an electron hits the surface of the TM,  a number of electrons corresponding to the yield for that energy is emitted from the surface according to \cite{grimcqg21}.
The role of electrons that elastically backscattered before reaching the TMs  above 1 keV  was also taken into account in FLUKA \cite{CSC}. 

\subsection{Electron and positron quantum backscattering}
The electron and positron quantum particle-wave duality cannot be neglected below 12 eV. As a result, quantomechanical effects such as quantum backscattering and diffraction are included in the simulation.

\begin{figure}
\centering
\resizebox{0.8\hsize}{!}{\includegraphics{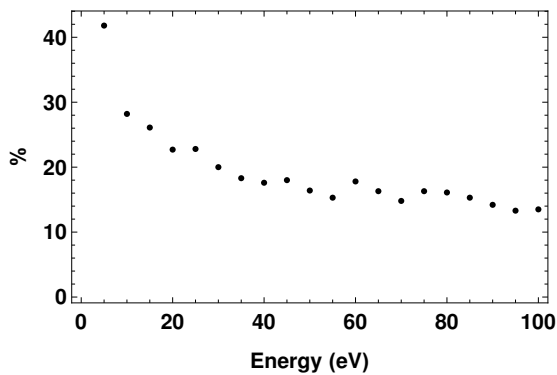}}
\caption{Quantum backscattering probability for electrons and positrons incident on a gold slab as a function of the energy.}\label{fig:qbs}
\end{figure}

The probability of backscattering is calculated by estimating the scattering amplitude. This calculation for gold was carried out in \cite{grimcqg21} by solving the Schr\"odinger equation for e$^-$ and e$^+$  propagating inside the lattice of a gold crystal. We found that the probability of these particles being backscattered strongly depends on their energy  and  decreases from about 42\%  at 5 eV to about 13\% at 100 eV, as shown in Figure \ref{fig:qbs}.


\section{FLUKA and LEI Monte Carlo simulation uncertainties}\label{sec6}
The Monte Carlo simulation results are affected by statistical and systematic uncertainties associated with the number of simulated events and the 
parameterizations of the physical processes. For each  run, about two million events were simulated.
The statistical errors were kept below 10\% and 5\% on the net and effective TM charging, respectively.
Further improving of these statistical uncertainties would have been meaningless because of the Monte Carlo intrinsic uncertainties.  
The comparison of the FLUKA Monte Carlo outcomes and beam experiment data discussed in \cite{fluka_uncertainty} has shown that the Monte Carlo outcomes are consistent with observations within 10\%.
The LEI uncertainties are discussed below for each process.

\subsection{Ionization}
\begin{figure}
\centering
\resizebox{0.8\hsize}{!}{\includegraphics{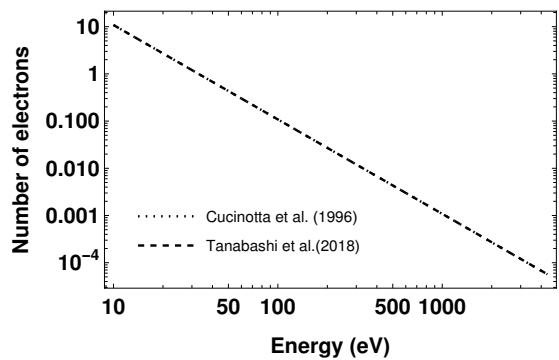}}
\caption{Comparison of the number of electrons produced by 2 MeV protons incident on a 150 nm gold slab according to \cite{cuci} and \cite{pdb}.
The two models overlap perfectly.}\label{fig:2mev}
\end{figure}

\begin{figure}
\centering
\resizebox{0.8\hsize}{!}{\includegraphics{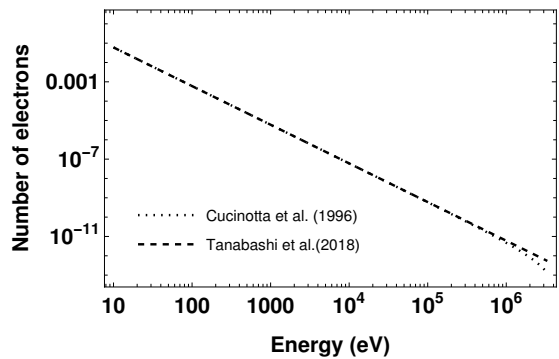}}
\caption{Same as Figure \ref{fig:2mev} for 1 GeV protons.}\label{fig:1gev}
\end{figure}

In Figures \ref{fig:2mev} and \ref{fig:1gev}, our parameterization of the number of secondary electrons produced by ionizing 2 MeV and 1 GeV protons propagating through a gold slab of 150 nm thickness is compared to \cite{pdb}.  The two parameterizations agree well: only a small deviation is observed above 100 keV
for 1 GeV incident protons. 

The parameterization for electron production by ionization reported in  \cite{cuci} was previously adopted by \cite{kk}, who 
compared the model to experimental data in lead and gold. They reported good agreement. 

The validation of the cross section adopted in LEI for  electron energy losses through ionization is discussed in \cite{sakata}. 
 They found that the uncertainty on the ionization cross section  above 1 keV, where both ionization and bremsstrahlung dominate,  appears to be a  few percent because data and model overlap within the data errors. No experimental data are available below 1 keV for  comparison in \cite{sakata}.


\subsection{Kinetic emission}
The electron yield for carbon, nitrogen, and oxygen nuclei incident on a gold slab inferred  from \cite{cfr} was compared with our calculations reported in \citet{grimcqg21} and implemented in LEI. We point out that \cite{cfr} did not use fully ionized incoming atoms, retaining  one electron, while in our calculations, the atoms were considered fully ionized. 
The overall difference of the yield was of 15\% for IIEE. Conversely, the electron yield associated with EIEE estimated according to \citet{schou80} exceeds the recent experimental work by \cite{cimino} by 30\%. 


\subsection{Quantum backscattering}
In \cite{grimcqg21} we calculated the probability of quantum backscattering from a slab of gold for incident electrons with energies lower than 100 eV.  Experimental  electron backscattering data below 100 eV are reported in \cite{QB}, who calculated the backscattering probability also for several elements, including gold. The probability of backscattering in gold appears to increase at low energies as in our calculations; in particular, based on their results, we estimated a backscattering probability of about 26\% at 50 eV. In fair agreement, we have found 20\% at 50 eV. An overall 10\% uncertainty on quantum backscattering yield was set as a result. 
Because low-energy electrons escape the gold-plated layers from electrodes toward the TMs and vice versa, these uncertainties 
constitute upper limits to those that actually affect the calculations of the net and effective charging.

\section{Lisa Pathfinder in-orbit test-mass charging and FLUKA/LEI simulations\label{sec7}}

The final geometry and material distribution of LISA S/C is not yet available. However, based on the preliminary design, it is plausible to assume that the material grammage of the LISA and LPF S/C will be similar. In order to set a range of reasonable values of the expected TM charging for LISA, we carried out new simulations with the FLUKA/LEI toolkit of the LPF TM charging at solar minimum, maximum, and during BR 2492 in order to compare these new simulation outcomes with the measurements carried out with LPF in April 2016.  
The cosmic-ray input fluxes are reported in Figures \ref{fig:p}-\ref{fig:moska}.  
The results of the simulations appear in Tables \ref{tab:maxmin}-\ref{tab:altra}. Table \ref{tab:maxmin} shows that at solar minimum and maximum, the main contribution to the TM net and effective charging
is given by  protons. Helium and heavy nuclei increase the net charging by a few percent and  the effective charging at solar minimum and maximum by about 50\%, even thought they constitute  just about 10\% of the cosmic-ray sample. 
This result arises because the large  production of electrons by ionization increases with the square of the nucleus charge.
On the other hand, we confirmed that a large number of  low-energy secondary electrons that escape TMs and electrodes contribute only little to the net TM charging. The GCR electrons, representing  only 1\% of the cosmic-ray sample, lower  the net charging  by about 15\% and increase the effecitve charging by approximately 10\%. 

In the simulations for BR 2492, during the declining phase of solar cycle 24, the contribution of nuclei with $Z>2$ to the net and effective charge is estimated from Table \ref{tab:maxmin} as an average between solar minimum and maximum conditions. As we recalled above, nucleus flux models at intermediate solar modulation conditions would  introduce  uncertainties larger  than the contribution of each of these nuclei to the net TM charging. We estimate that nuclei account for net and effective charging of 2-4\% and 10-20\% of the total, respectively.
The slightly negative charging associated with heavy nuclei is due to statistical fluctuations in the low-energy electron production.  

The simulation results reported in Tables \ref{tab:2492} and \ref{tab:altra} for the two TMs can be compared with the TM net and effective charging measurements  carried out with LPF in April 2016. They are listed in Table \ref{tab:meas}. Our simulations  agree  excellently with observations when the heavy nuclei contribution is taken into account and within the measurements and Monte Carlo uncertainties discussed in the previous sections. 
Moreover, the contribution of the low-energy electromagnetic physics is highly important to estimate the LPF TM effective charging. Our work allows us to bridge the gap between Monte Carlo simulations and observations for LISA.  


\begin{table*}
\renewcommand*{\arraystretch}{1.3}
\caption{Average FLUKA/LEI LPF TM charging at solar minimum and solar maximum for LISA.}\label{tab:maxmin}
\centering
\begin{tabular}{c|cc|cc}
\hline\hline
&\multicolumn{2}{c|}{Solar Minimum}&\multicolumn{2}{c}{Solar Maximum}\\
& Net charging & Effective charging rate & Net charging & Effective charging rate\\
Primary particle & (e$^{+}$ s$^{-1}$)  & (s$^{-1}$) & (e$^{+}$ s$^{-1}$) & (s$^{-1}$)\\
\hline
Protons &  50.4 & 937.9&  5.30 & 316.6\\
$^4$He & 5.8& 182.5 & 0.28& 108.9 \\
$^3$He & 2.9& 90.5 & 0.07& 54.0\\
Nitrogen & -0.1 & 11.3 & 0.02 & 6.4\\
Carbon & 0.4 & 31.9 & 0.08 & 19.6 \\
Oxygen & 0.5 & 80.6 & 0.13 & 34.0\\
Iron & 0.1 & 27.5 & -0.04 & 26.5\\
Electrons & -7.8 & 167.8 & -1.2 & 53.6 \\
Total & 52.2 & 1530.0 & 4.64 &  619.6\\
\hline
\end{tabular}
\end{table*}


\begin{table}
\renewcommand*{\arraystretch}{1.3}
\caption{FLUKA/LEI LPF first TM charging during the BR 2492.}\label{tab:2492}
\centering
\begin{tabular}{ccc}
\hline\hline
& Net charging & Effective charging rate\\
Primary particle & (e$^{+}$ s$^{-1}$) & (s$^{-1}$)\\
\hline
Protons &  21.35 & 611.8\\
Helium  &  3.9  & 201.2\\
Electrons& -4.1 & 75.5\\
Total   &  21.15 & 888.5\\
\hline
\end{tabular}
\end{table}

\begin{table}
\renewcommand*{\arraystretch}{1.3}
\caption{Same as Table \ref{tab:2492} for the second LPF TM.}
\label{tab:altra}
\centering
\begin{tabular}{ccc}
\hline\hline
& Net charging & Effective charging rate\\
Primary particle & (e$^{+}$ s$^{-1}$) & (s$^{-1}$)\\
\hline
Protons &  20.5 & 673.6 \\
Helium & 5.8 & 161.9 \\
Electrons  & -4.3 & 98.0 \\
Total & 22.0 & 933.5\\
\hline
\end{tabular}

\end{table}


During the LPF mission, no SEP events were observed. In order to estimate the  possible  effects on LISA,  the charging of the LPF TMs was calculated for  the three  SEP events of different intensities dated February 23, 1956, December 13, 2006, and December 14, 2006, as reported in Section \ref{sec3}. 
The TM charging  at the onset and peak of each event is reported in Table \ref{tab:SEP}. 
The charging of the TMs during the evolution of these SEP events increases by several orders of magnitude with respect to that  induced by GCRs.
These estimates will be useful to optimize the TM discharging  process during the LISA operations. 

\begin{table}
\renewcommand*{\arraystretch}{1.3}
\caption{Average FLUKA/LEI LPF TM charging during the evolution of SEP events.}\label{tab:SEP}
\centering
\begin{tabular}{ccc}
\hline\hline
\multicolumn{3}{c}{02/23/1956}\\
\hline
& Net charging & Effective charging rate\\
& (e$^{+}$ s$^{-1}$) & (s$^{-1}$)\\
\hline
Onset & 8870 & 73730\\
Peak & $1.3 \times 10^8$ & $1.3 \times 10^8$\\
\hline
\multicolumn{3}{c}{12/13/2006}\\
\hline
& Net charging & Effective charging rate\\
& (e$^{+}$ s$^{-1}$) & (s$^{-1}$)\\
Onset & 2425 & 5695\\
Peak & 1123 & 2360\\
\hline
\multicolumn{3}{c}{12/14/2006}\\
\hline
& Net charging & Effective charging rate\\
& (e$^{+}$ s$^{-1}$) & (s$^{-1}$)\\
\hline
Peak & 88.7 & 141.3\\
Decay & 3.3 & 4.9\\
\hline
\end{tabular}
\end{table}

\section{Conclusions}\label{sec8}
High-energy particles of solar and Galactic origin will  charge the TMs of the future interferometers for gravitational wave detection in space. 
Monte Carlo simulations carried out with Geant4 and FLUKA toolkits before the LPF launch allowed us to estimate the net and effective charging 
for the first part of the mission (end of 2015 to the beginning of 2016) on the basis of the expected solar activity. Measurements carried out with LPF in April 2016
were compared to simulations. The observed net charging of about 23-25 positive charges per second agreed with expectations, while the estimated charging noise was three to four times lower than in-orbit observations. This mismatch probably arises because low-energy electromagnetic 
processes below 1 keV are missing in FLUKA. 

The results of new LPF TM charging simulations 
are reported in this work. A dedicated Monte Carlo (LEI) was written to include low-energy electron propagation and quantum backscattering down to a few electronvolt. These new simulations appear to agree with the observations within the errors of the Monte Carlo simulation and measurements. We also considered solar minimum and solar maximum conditions and SEP events for LPF as reasonable predictions for the TM charging of LISA. The details of the LISA S/C geometry are not yet available, even though the  amount of material surrounding the TMs is expected to be similar to that of LPF. The net (effective) charging will probably vary between a few (hundreds) and about 50 (thousand) charges per second due to GCRs. The charging is estimated to increase by several orders of magnitude during SEP events.
The evolution of three SEP events with a fluence ranging between 10$^6$ and 10$^9$ protons cm$^{-2}$ was considered. 
The net and effective charging induced by cosmic rays differ by a few  orders of magnitude, while during SEP events,
the net and effective charging are similar because the SEP energy spectra present higher spectral indices above hundreds of MeV  with respect to GCRs. As a result, the majority of particles stop in the material surrounding the TMs or in the TMs with a minor secondary particle production due to cascading, which increases the effective charging more than the net charging. The LISA TM charging estimates will allow us to control the TM charging process, to optimize the discharging during the mission operations, and to evaluate the role of the charging noise with respect to the total mission noise budget.  


\begin{acknowledgements}
Part of this work was funded under the European Space  Agency project AO-1-10081 - TEST MASS CHARGING TOOLKIT AND LPF LESSONS LEARNED.
\end{acknowledgements}

%
   \bibliographystyle{aa} 
   \bibliography{biba25Apr22} 
%

\end{document}